\documentclass[sigconf,nonacm]{acmart}

\AtBeginDocument{%
  }

\setcopyright{acmcopyright}
\copyrightyear{2023}
\acmYear{2023}

\usepackage{listings}
   
\begin{document}

\hyphenation{Exa-MPI}

\title{MPI Implementation Profiling for Better Application Performance}

\author{Riley Shipley}
\orcid{0000-0002-7486-7542}
\affiliation{%
  \institution{Tennessee Technological University} 
  \streetaddress{1 William L Jones Dr}
  \city{Cookeville}
\state{TN}
\country{USA}
}
\email{rpshipley@tntech.edu}

\author{William Garrett Hooten}
\affiliation{%
    \institution {University of Tennessee, Chattanooga}
\city{Chattanooga}
\state{TN}
\country{USA}
}
    
\author{David Boehme}
\orcid{0000-0002-4159-1519}
\affiliation{%
    \institution{Lawrence Livermore National Lab}
    \city{Livermore}
    \state{CA}
    \country{USA}
}
\email{boehme3@llnl.gov}

\author{Derek Schafer}
\orcid{0000-0001-8438-5144}
\affiliation{%
    \institution{University of New Mexico} 
  \city{Albuquerque}
  \state{NM}
  \country{USA}
}
\email{dschafer1@unm.edu}

\author{Anthony Skjellum}
\orcid{0000-0001-5252-6600}
\affiliation{%
  \institution{Tennessee Technological University} 
  \streetaddress{1 William L Jones Dr}
  \city{Cookeville}
  \state{TN}
  \country{USA}
}
\email{askjellum@tntech.edu}

\author{Olga Pearce}
\affiliation{%
    \institution{Lawrence Livermore National Lab}
    \city{Livermore}
    \state{CA}
    \country{USA}
}
\email{pearce8@llnl.gov}

\renewcommand{\shortauthors}{Shipley, Hooten et al.}

\begin{abstract}
While application profiling has been a mainstay in the HPC community for years, profiling of MPI and other communication middleware has not received the same degree of exploration. 
This paper adds to the discussion of MPI profiling, contributing two general-purpose profiling methods as well as practical applications of these methods to an existing implementation.
The ability to detect performance defects in MPI codes using these methods increases the potential of further research and development in communication optimization.
\end{abstract}
\maketitle


\section{Introduction}
Application performance is not solely dependent on the capabilities of the hardware upon which it is run but also how well the application utilizes the hardware resources it is allocated.
Instead of focusing directly on application optimizations, this paper emphasizes understanding the characteristics and performance behavior of MPI implementations to positively impact overall application performance.

To this end, we propose two profiling methods that can make these goals achievable for both parties.
The first method uses the performance of a baseline MPI implementation as a comparison point to assess the performance of an implementation of interest. This method does not require access to the MPI implementation's source code, so it can be used by application developers to decide which MPI implementation is best for their application on a given platform, and also by implementers to inform them of how well their implementation performs relative to other implementations.
The second method is a finer-grain approach focused on revealing problematic behaviors or patterns within the MPI implementation by marking and profiling regions of interest within the MPI implementation's source code. An application or benchmark using the profiled implementation is then run, creating a visual timeline of profiled events in order to reveal any problematic behaviors.

 The remainder of this paper discusses the libraries and applications used throughout (Sec.~\ref{libraries}) and introduces comparison\-based profiling and timeline profiling in Sections ~\ref{comparison-based} and ~\ref{timeline}, respectively.

\section{Libraries \& Applications}
\label{libraries}
This section covers all the key details of the software utilized for this paper. 
While this section provides details about the specific MPI implementations, profiling tool, and application used for the demonstrations discussed in this paper, all of these could be substituted with other libraries and applications with similar capabilities.

\subsection{ExaMPI}
\label{ExaMPI}
ExaMPI was created by Tennessee Technological University in collaboration with other universities and laboratories with the intent to be an easily modifiable MPI testbed \cite{ExaMPI}. 
Unlike complete MPI implementations, which strive to implement all MPI features regardless of their current utility, ExaMPI takes a different approach by focusing development efforts on portions of the MPI standard that are still highly relevant in today's computing ecosystem; more features are implemented as the need arises. 
Using this design, ExaMPI provides a simpler environment for MPI implementation experimentation by not having to maintain legacy code or older MPI Standard interfaces. 

Internally, ExaMPI has two more significant differences from other MPI implementations: it is written in C++, and features a separate, per-process devoted communication thread to support overlap between communication and computation.
Because it is written in modern C++, ExaMPI enables developers to take advantage of many memory safety, thread safety, and productivity features not native to C. 
This built-in functionality helps ExaMPI remain understandable and readable.

The strong progress engine is the second significant feature of ExaMPI. 
By dedicating a separate thread solely to the purpose of progressing communication requests, ExaMPI allows for direct overlap of computation and communication, in contrast to many other implementations, which only work on communication when no computation is being done.
Though the strong progress is a nice addition, the main draw of ExaMPI for this study was the ease of understanding and modification its research-friendly design promotes \cite{ExaMPI}.

\subsection{Caliper}
\label{Caliper}

Caliper~\cite{Caliper} is a performance instrumentation and profiling library for HPC codes. 
It is primarily designed around an instrumentation API that lets developers mark regions of interest in the source code. 
In large HPC codes, this manual instrumentation approach ensures precise control over the instrumented program locations and the overall amount of collected data. 
Developers can thus analyze program performance in terms of high-level program abstractions like kernels or phases. 
In contrast, automated approaches relying on compiler-generated symbol names often produce obscure associations, in particular for modern C++ codes.

Typically, Caliper's source code region annotations can remain in the code throughout its life cycle. 
Moreover, region annotations on different layers in the software stack (for example, a HPC application and a library) are combined into a single context tree. 
Performance measurements can be enabled at runtime. 
Caliper provides many built-in performance measurement recipes, covering a wide range of analysis use cases from always-on profiling to detailed event tracing. 
A variety of different output formats are available, including human-readable text tables, machine-readable profiles that can be read with the Hatchet library, and detailed event traces. 
These traces can be converted for visualization with the Chromium trace viewer ~\cite{chromium-trace}, Perfetto~\cite{Perfetto}, NSight~\cite{Nsight}, or Vampir~\cite{vampir}.

\subsection{COMB}
\label{COMB}

COMB~\cite{Comb} is a communication performance benchmark specifically designed to investigate tradeoffs in implementing communication patterns on heterogeneous, accelerator-based systems. 
To that end, COMB enables exploration of a large design space over different communication patterns (e.g., blocking or non-blocking communication), execution strategies (e.g., OpenMP, CUDA, or serial CPU execution), and memory spaces for staging buffers (e.g., GPU-memory buffers, host-memory buffers, pinned buffers, etc.). 
COMB is fully instrumented with Caliper source code annotations for detailed analysis of the observed performance behavior.

\section{Comparison-based Profiling} \label{comparison-based}
In this section, we discuss the comparison-based profiling methodology, including how to use it and applications for it as well as demonstrating the method on ExaMPI and discussing the outcomes of applying this method. 
\subsection{Methodology}
As the name suggests, comparison-based profiling compares the performance of one MPI implementation to another by running two identical sets of tests, each using a different implementation.
The goal of this approach is to determine the performance of MPI procedure calls in the experimental implementation and the baseline implementation in order to discover areas that perform the worst in the experimental library as a starting point for optimization efforts.
It is important to note that the baseline implementation is not expected to be perfectly optimized; such an implementation likely does not exist.
Rather than being perfect, the baseline only needs to perform well in the specific areas of interest to the implementer, since that should be the focus of comparison.

In addition to the two MPI implementations, this method also requires a performance profiling tool that can automatically time MPI procedure calls and an application to run with each implementation in order to generate performance data.
There are many tools that can automatically time MPI calls: Tau~\cite{Tau}, Vampir~\cite{vampir}, Caliper~\cite{Caliper}, and many other profiling tools have this capability.
For our experiment, we chose to use Caliper because it allowed us to more easily use Hatchet~\cite{Hatchet} for analysis of the application comparison data, which will be discussed later on.
Similarly, any application or benchmark that uses MPI can be used for this method. However, some care should be taken when choosing an application since each one uses MPI in its own way and will thus reveal different performance characteristics.

With the desired application and profiling tool chosen, the next step is to build and run the application using each MPI implementation. 
The configuration of the application is entirely up to the experimenter. However we recommend disabling MPI+X features of the application (unless they are the direct focus of the experiment) and running with each implementation many times to obtain more precise feedback. 

Once the runs have completed, the final steps are to aggregate the collected call completion times for each implementation and create a comparison ratio for each MPI procedure call.
Some care should be taken when aggregating the data according to the intent with which the methodology is applied: averages may be appropriate in many cases, but there are many aspects of MPI that may be more appropriately measured in terms of maximums, minimums, or overall variance.
Finally, the two aggregated data sets are used to create a ratio representing the speedup (or lack thereof) of the experimental implementation relative to the baseline implementation.
To do this, each aggregate call time for the baseline implementation is divided by the corresponding call time for the experimental implementation.
Comparing the two in this way creates a ratio that is easily interpreted: comparison values greater than one show the experimental implementation has superior performance, while values less than one represent inferior performance and values approximating one represent roughly equal performance.

\subsection{Experimentation}

In the inaugural application of this method, the authors sought to evaluate the overall performance of ExaMPI on Lawrence Livermore National Laboratory's Lassen cluster.
COMB~\cite{Comb} was chosen for this method because it is a general-purpose communication benchmark and it comes packaged with build scripts for Lassen, which aligned nicely with the goals of the experiment since we would be able to use a pre-tuned application rather than try to tune it ourselves.
COMB also has its own set of Caliper profiling regions, which provided additional context as to what the MPI procedure calls are being used for and how MPI performance might factor into overall performance.
Within COMB, there are a variety of testing configurations provided, but for this paper, we used a modified version of the \textit{focused\_mpi\_type\_tests} configuration script to set up COMB.
The two important changes made to this script were the addition of Caliper profiling, in order to collect timing data, and the repetition of the test, so that many results could be collected on one job allocation so changes in network topology and latency were minimal.

While any MPI implementation could have been used, implementations that were already part of Lassen's software stack were preferred to avoid adding the complexity of building an implementation that was not already installed, as building and using an MPI implementation for the first time on Lassen would likely have under-performed relative to one which has been used and fine-tuned prior to the experiment.
Of the few remaining options, Spectrum MPI \cite{Spectrum} was chosen because it comes from the same company (IBM) that manufactured Lassen's Power9 processors \cite{Lassen}.
The reasoning behind this was that Spectrum, as an IBM product, was more likely to have specific optimizations for the IBM system than implementations from other vendors or organizations.

\begin{figure}
\centering
\includegraphics[width=0.9\linewidth,trim={1em 31em 2em 0},clip]{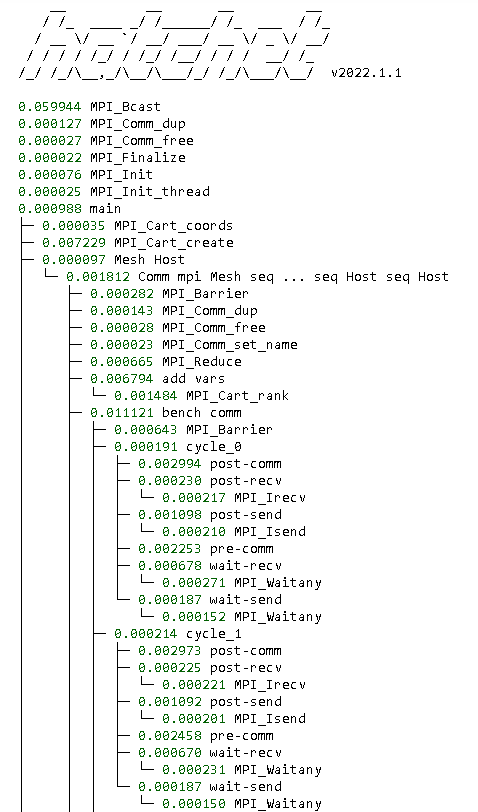}
\caption{Top of a Hatchet tree containing average completion times of ExaMPI at the start of the experiment}
\label{example-tree}
\end{figure}

After everything was compiled and configured correctly, the modified \textit{focused\_mpi\_type\_tests} configuration of COMB was run on 27 processes distributed over 7 nodes, each of which were exclusive to the application during its run to minimize the impact of network interactions.
Though Lassen nodes have up to 40 processors available per node \cite{Lassen}, the MPI processes were distributed across multiple nodes to mimic a more realistic network topology in which processes communicate with both on- and off-node peers. 
After all runs had completed, Hatchet~\cite{Hatchet,hatchet_protools_2020} was used to aggregate and analyze all of the individual timing files. 
Hatchet is a Python analysis tool created by LLNL that transforms data produced by Caliper, HPCToolkit, and other profilers and data analysis tools into condensed, easily readable graph-like data structures, called GraphFrames (based on the pandas DataFrame~\cite{pandas-paper}).
From here, data from individual files (or rather their corresponding GraphFrames) can be transformed and aggregated using many of the same operations supported by DataFrames.
The main difference between GraphFrames and DataFrames, and the reason Hatchet was used for this analysis, is that GraphFrames are able to maintain a hierarchical structure throughout analysis, as shown in Figure~\ref{example-tree}.

Using Hatchet, all of the individual calls for the many application runs using each implementation were aggregated using a mean function before the corresponding calls for each tree were divided by one another.
Rather than using a DataFrame function or writing a new function to accomplish this, Hatchet provides the capability to perform simple arithmetic with GraphFrames, making a potentially complex task extremely simple.

\begin{figure}
\centering
\includegraphics[width=0.9\linewidth,trim={40 14.5em 0 30.2em},clip]{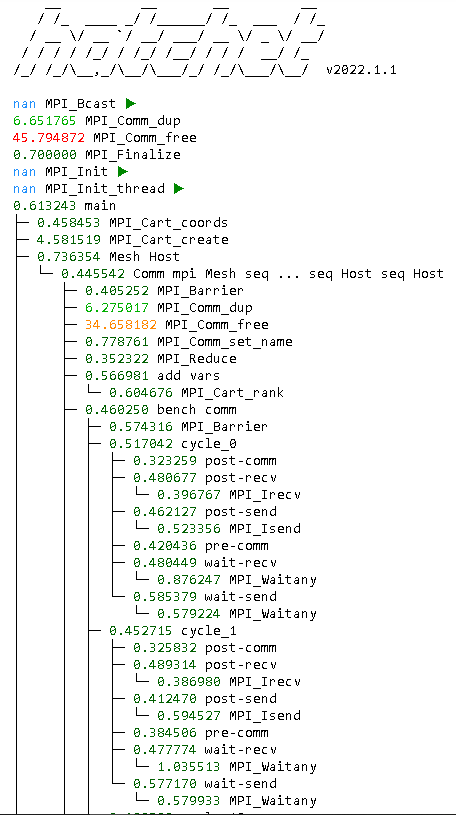}
\caption{Portion of a Hatchet tree showing ExaMPI's lower performance relative to Spectrum on both communication and computation tasks}
\label{comparison-tree}
\end{figure}

In the ExaMPI-Spectrum comparison in Figure~\ref{comparison-tree}, ExaMPI was originally much slower than Spectrum across almost every MPI procedure call, shown by the overwhelming number of values less than one.
After looking more closely at the individual regions in the comparison tree, we realized that it is not only MPI functions which behave this way, but also computation-only regions in COMB, in which no MPI functions are called.

\begin{figure}
\centering
\includegraphics[width=0.9\linewidth,trim={40 14em 0 26.5em},clip]{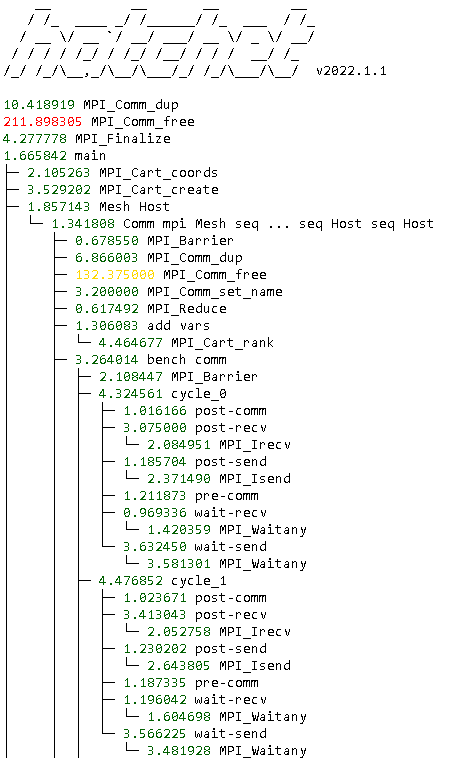}
\caption{Portion of a Hatchet tree showing the improved ExaMPI's performance relative to Spectrum}
\label{new-exa-tree}
\end{figure}

After seeing this result, it became clear that something was not right in ExaMPI, and pursuing this behavior revealed that there was a core scheduling issue, specifically that some of the cores allocated for a given application run were being over-subscribed by the number of processes assigned to them, while others were not used at all.
When the issue was fixed, the experiment was run again to determine the impact of the changes, resulting in a more optimistic comparison tree shown in Figure \ref{new-exa-tree}.
The new results showed that ExaMPI completed MPI procedure calls faster than Spectrum in most cases, and that computation regions (such as \texttt{pre-comm} and \texttt{post-comm}) took approximately the same amount of time for both implementations, confirming that the core scheduling was the cause for the mismatch in computation times.

\subsection {Results}

\begin{figure}
    \centering
    \includegraphics[width=\linewidth,trim={5 5 5 5},clip]{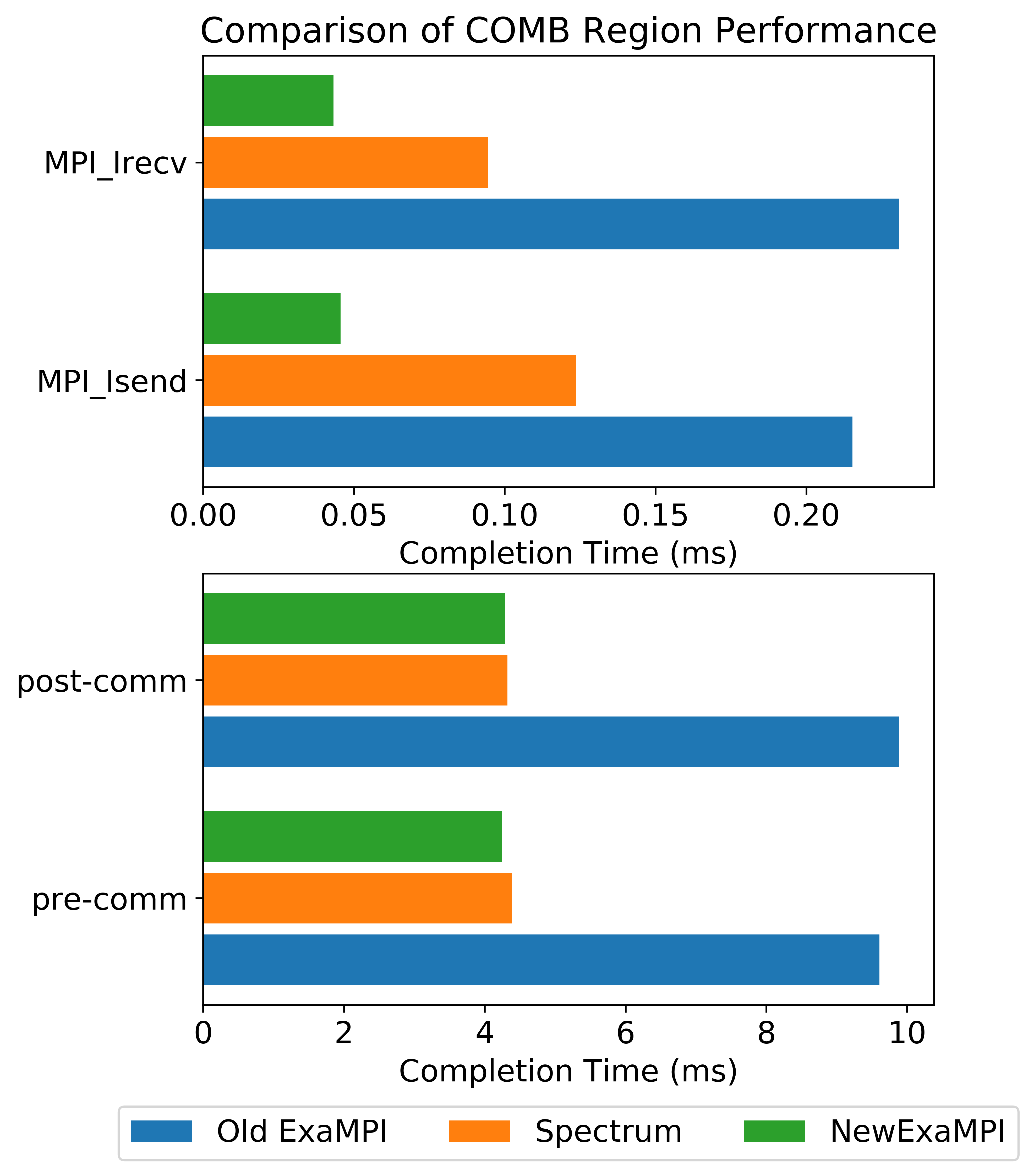}
    \caption{Comparison of ExaMPI before and after core scheduling changes to Spectrum}
    \label{fig:comparison-summary}
\end{figure}

After adjusting ExaMPI as described above, COMB was run again using all three implementations to evaluate the effectiveness of the changes made to ExaMPI, and thus evaluate the performance gained by applying this method.
The data in Figure \ref{fig:comparison-summary} represents the average of a set 50 runs of COMB each for the old ExaMPI (prior to core scheduling improvement), Spectrum MPI, and the improved ExaMPI.

\begin{figure}
    \centering
    \includegraphics[width=0.9\linewidth,trim={5 5 5 5},clip]{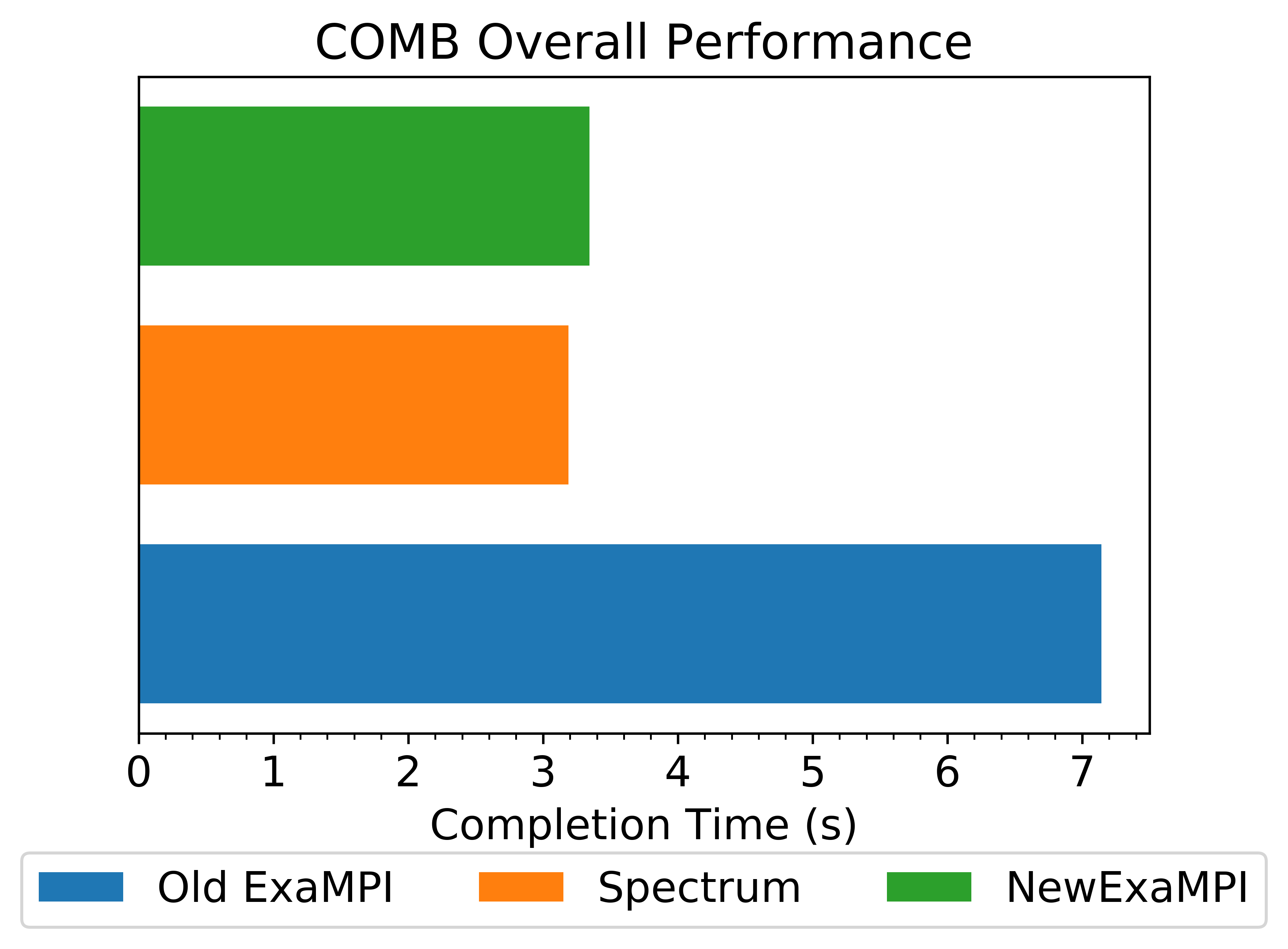}
    \caption{Comparison of COMB completion times between all 3 implementations}
    \label{fig:comb-overall-time}
\end{figure}

The updated ExaMPI performs similarly to what was expected prior to using this method: compute regions perform similarly to Spectrum, because the MPI implementation used should have little effect there, and MPI calls perform better than Spectrum on average, resulting in an average speedup of 3.58$\times$ across all MPI procedure calls used in COMB.
The authors expected ExaMPI to be faster than Spectrum for MPI procedure calls because it features an active progress thread that progresses messages in the background, unlike many other implementations, including Spectrum, that may only work to fulfill requests when they are forced to by another MPI procedure call \cite{ExaMPI-2}.
Overall, the average run time of COMB was reduced by 44.66\% by replacing the old ExaMPI implementation with the one improved by the comparison-based profiling method, as shown in \ref{fig:comb-overall-time}.

For MPI researchers and implementers, comparison-based profiling can provide an easy way to prioritize which areas of a communication interface to work on next, or a way to quickly evaluate the results of a change that has been implemented.
The versatility of this tool helps application developers tune many facets of MPI use within their applications, such as testing various MPI implementations to know which one is favorable, as well as testing different patterns of communication using the same implementation.
Overall, comparison-based profiling is a widely accessible starting point for anyone wanting to reduce communication overhead that produces easily understandable insights into an MPI implementation's characteristics without needing access to its source code.

\section{Timeline Profiling} \label{timeline}
In this section, we will discuss the timeline profiling methodology, including how to use it and applications for it as well as demonstrating the method on ExaMPI and discussing the outcomes of applying this method. 
\subsection{Methodology}
Timeline profiling consists of collecting data from inside of an MPI implementation over the course of an application run and observing the resulting trace to find opportunities for optimization.
Because the traces produced can be quite large, it is best to have a focus or area of interest going into this method rather than just following the timeline as the latter can be overwhelming and somewhat confusing.
In general, areas with poor performance discovered through other means - such as the comparison-based method above - make an ideal place to start applying timeline profiling.

Once the areas of interest for a particular implementation have been decided, a tracing tool such as Caliper or Tau is used to collect timing data by adding new timing regions within the MPI implementation of interest.
In the case of Caliper, a region can be added to a program's timing profile simply by wrapping the region in Caliper annotations as shown in Figure~\ref{fig:cali-region}.
\begin{figure}
\centering
\begin{lstlisting}[language=C++,frame=single,breaklines=true,keywordstyle=\textbf]
#include <caliper/cali.h>

int main(int argc, char *argv[])
{
    cali_begin_region("main");
    // Code here
    cali_end_region("main");
}
\end{lstlisting}
\caption{Creating a profiled code region with Caliper.}
\label{fig:cali-region}
\end{figure}

From here, an application or benchmark of choice is run one time using the annotated implementation and the resulting data is fed back to the tracing tool to be converted into a timeline format before it is observed in the appropriate timeline viewer (NVIDIA NSight~\cite{Nsight} or Chromium~\cite{chromium-trace} for Caliper, Jumpshot~\cite{Jumpshot} for Tau, etc.).

While there is no one-size-fits-all method for analyzing the timeline, we suggest the following activities as part of this method: 
\begin{itemize}
    \item Checking for large waits in synchronizing functions, specifically collective operations (e.g.,  barriers and reductions)
    \item Thoroughly analyzing critical sections of any parallel regions for delays due to thread contention
    \item Investigating regions that are irregular in duration relative to other occurrences of the same code region
    \item Analyzing large gaps between profiled regions
\end{itemize}
Once an area of interest has been identified on the timeline, it is up to the observer to carefully analyze the events captured to locate potential sources of the problematic behaviors identified, at which point it is often helpful to compare the timeline to the source code of the application and the MPI implementation to pinpoint the issue.

\subsection{Experimentation}

In the case of ExaMPI, we were not necessarily focused on any singular area of interest but sought to further our exploration of ExaMPI's performance characteristics on Lassen as a whole.
Because we knew profiling all of ExaMPI would be a sizeable undertaking, we chose to integrate Caliper profiling into ExaMPI so Caliper could be used in future projects without duplication of effort.
Rather than adding regions directly, we took advantage of ExaMPI's existing runtime configuration capabilities to add flexible levels of profiling.
Functions within ExaMPI were divided into four separate categories that can each be turned on or off at runtime to limit profiling overhead and lower the size of the trace files Caliper creates.

\begin{figure}
    \centering
    \includegraphics[width=0.9\linewidth]{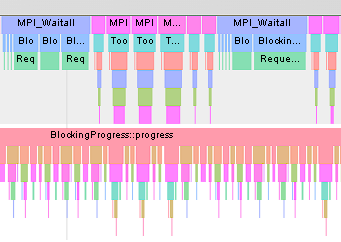}
    \caption{A macro-scale view of an example ExaMPI timeline}
    \label{fig:lock-full-trace}
\end{figure}

After adding the flexible profiling regions to ExaMPI, COMB was run with ExaMPI once again to produce trace files, which Caliper then converted to the Chromium trace format \cite{chromium-trace}.
The end result, shown in Figure \ref{fig:lock-full-trace}, is a color-coded, interactive, and searchable timeline of all profiled regions that occurred during application execution.

\begin{figure}
    \centering
    \includegraphics[width=0.9\linewidth,trim={0 3em 0 0},clip]{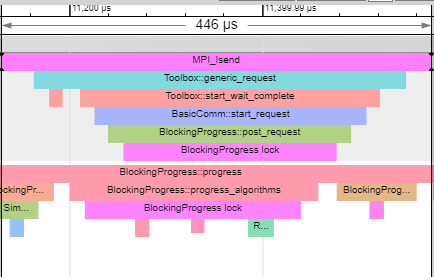}
    \caption{A portion of ExaMPI's execution timeline showing lock contention between user and progress threads}
    \label{fig:lock-contention-before}
\end{figure}

Once thorough investigation of the timeline began, one trend quickly emerged from the timeline, shown in detail in Figure \ref{fig:lock-contention-before}.
The figure shows both threads of ExaMPI, the user thread (top) and the progress thread (bottom), entering the same region labeled ''BlockingProgress lock.''
This region represents a mutex lock associated with a request processing queue used by both threads, on which the user thread enqueued events that were read by the progress queue, which then completed the actions necessary to satisfy each request before it was removed from the queue.
As shown in the figure, the two threads could and did indeed contend for the lock,  which resulted in a prolonged waiting period for the thread that arrived latest.

After realizing this, an additional incoming request queue was added so the user thread would never be waiting on the progress thread to relinquish the processing queue lock in order to maximize potential for strong progress.
The progress thread now moves the incoming queue requests to its own internal queue for processing, rather than using the same queue as the user thread.

\begin{figure}
    \centering
    \includegraphics[width=0.9\linewidth,trim={0 0 0 0},clip]{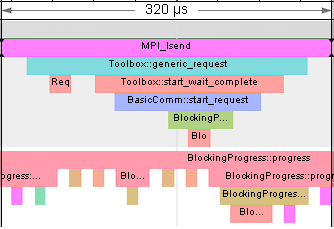}
    \caption{A portion of ExaMPI's execution timeline showing resolution of lock contention between user and progress threads}
    \label{fig:lock-contention-after}
\end{figure}

After implementing the second queue, COMB was rerun with the improved ExaMPI to double-check that the changes reduced the amount of time spent waiting for locks, which is clearly shown to be the case in Figure \ref{fig:lock-contention-after}.
In this timeline, the lock regions are no longer overlapping with one another because they rarely contend for the same lock, and as a result are much smaller due to the lack of contention between threads.
In addition to these per-MPI-process (rank) savings gained by keeping the user thread and the progress from facing contention with one another, the optimization of adding a second queue also greatly improved performance across the entire application, which is discussed further in the next section.

\subsection{Results} \label{timeline-results}
After adding the second queue described above, COMB was run again with both versions of ExaMPI to discern the amount of impact the lock contention (or lack thereof) had on the performance of the overall application.

\begin{figure}
    \centering
    \includegraphics[width=0.9\linewidth,trim={10 10 10 10}]{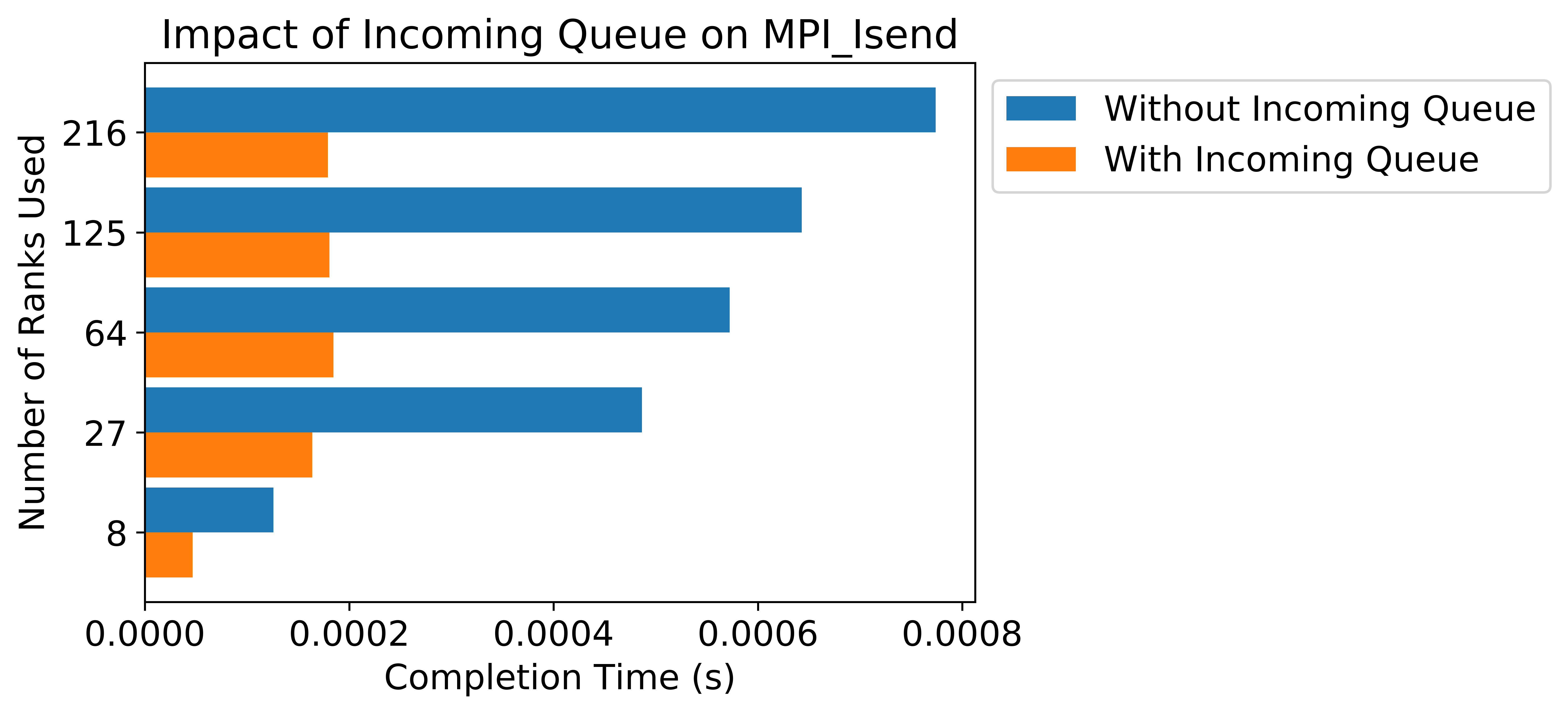}
    \caption{By adding a separate queue for incoming messages, ExaMPI was able to greatly reduce the amount of time spent in \texttt{MPI\_Isend}.}
    \label{fig:message-queue-isend}
\end{figure}

\begin{figure}
    \centering
    \includegraphics[width=0.9\linewidth,trim={10 10 10 10}]{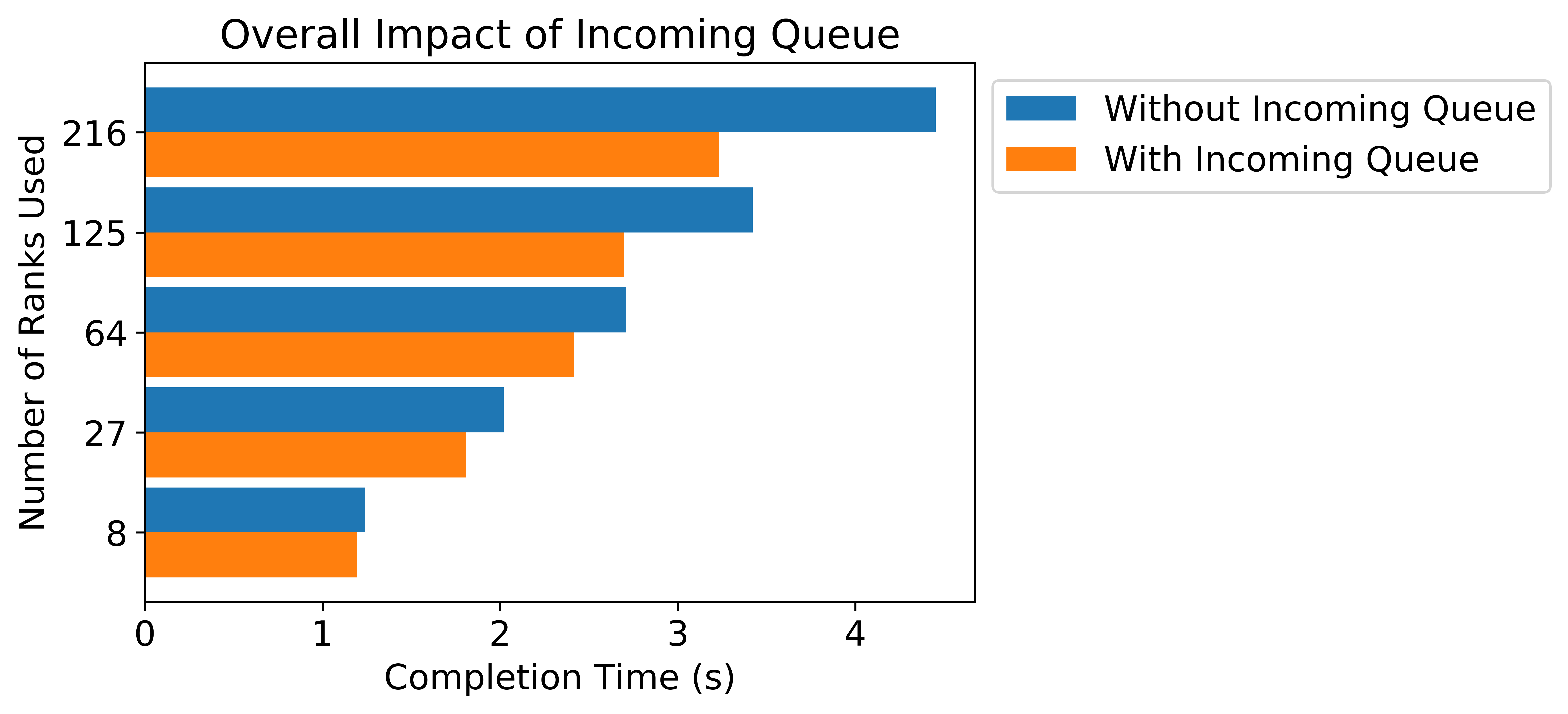}
    \caption{Impact of incoming message queue on overall application performance.}
    \label{fig:message-queue-overall}
\end{figure}

Figure \ref{fig:message-queue-isend} compares the time to completion for \texttt{MPI\_Isend} without the incoming queue to the time to completion for the revised version across of range of problem sizes.
Without the second request queue, the average completion time for \texttt{MPI\_Isend} grows with the number of MPI processes (ranks) used for the COMB run.
The presence of more MPI processes results in a larger queue of requests that need to be processed by the progress thread, which then causes the progress thread to hold the lock for the shared queue longer, leading to an increase in the average time to completion for \texttt{MPI\_Isend}.\\
This is also why the average \texttt{MPI\_Isend} completion time remains nearly constant when the additional queue is used.
In the improved model, the progress queue does not process requests while it holds the lock for the user request queue, thus the increasing number of pending requests caused by the increase in the number of MPI processes  does not impact the user thread and by extension does not affect \texttt{MPI\_Isend}.
COMB's overall performance also seems to have benefited from the ExaMPI improvements for all number of MPI processes that were tested, as shown in Figure \ref{fig:message-queue-overall}.

\section{Conclusions}
\label{sec:conclusions}
While application profiling and optimization methods have been thoroughly discussed, MPI profiling and optimization continues to provide many opportunities for innovation and progress. 

This paper proposes two methods for profiling MPI libraries, each with a specific function.
Comparison-based profiling captures data from two different MPI implementations used over several runs of an application or benchmark, which is then aggregated for each implementation.
Following this, the aggregated values are compared against one another by dividing the values of one (the ''experimental'' implementation) by the values of the other.
This produces a  single number that indicates which implementation performs better than the other, enabling users to quickly locate areas of the experimental implementation with the worst performance.

The second method, timeline profiling, requires a bit more manual work to set up, but can be applied repeatedly once the initial work is done.
Timeline profiling leverages the one-time addition of Caliper region markers to an MPI implementation's source code to provide the capability to generate fully interactive timelines of MPI regions over the course of any run of any application.
Carefully analyzing an application timeline can be especially useful as a detailed look into specific problem areas.

To demonstrate these methods, both were applied to ExaMPI using COMB as the test application, demonstrating that the methods proposed in this paper are capable of uncovering behaviors with significant impacts on the performance of communication libraries such as MPI, and by extension all of the applications that use MPI.
Thus, we encourage MPI users, implementers and developers to consider applying these profiling methods to continue optimizing past the application layer. 

\section{Future Work}\label{sec:future-work}
While in the process of researching this paper, the authors came upon a number of other interesting topics.
The first of these is co-profiling, or the profiling of both MPI and the application at the same time to gain application-specific insights.
The goal of this activity is to tune aspects of an application's MPI usage such as message size, timing of MPI procedure calls and so on to gain performance from the MPI implementation and ultimately the application.

Another is the use of message tracing in the context of timeline visualizations of the kind presented here.
Having knowledge of the exact paths messages take may lead to new insights on how to better structure an ideal MPI implementation for maximum performance.

Finally, we also plan to continue using and applying the methods seen here along with new benchmarks and applications in order to make further improvements to ExaMPI.

\section*{Acknowledgments}
\sloppy
This work was performed with partial support from the National Science
Foundation under Grants Nos.~CCF-1918987, 
CCF-1822191, CCF-1821431, OAC-1923980, OAC-1549812, OAC-1925603, OAC-2201497, and CCF-2151020; the U.S. Department of Energy's National Nuclear Security Administration (NNSA) under the Predictive Science Academic Alliance Program (PSAAP-III), Award DE-NA0003966; and under the auspices of the U.S. Department of Energy by Lawrence Livermore National Laboratory under Contract DE-AC52-07NA27344. LLNL-CONF-852648.

Any opinions, findings, and conclusions or recommendations expressed in this material are those of the authors and do not necessarily reflect the views of the 
National Science Foundation, the U.S.\hbox{} Department of Energy's National Nuclear Security Administration  or the Lawrence Livermore National Laboratory.

\bibliographystyle{ieeetr}
\bibliography{refs}

\vspace{12pt}

\end{document}